\documentclass[fleqn,12pt,twoside]{article}
\usepackage{espcrc1}

\newcommand{\cd}{\makebox[0.08cm]{$\cdot$}}

\newcommand{\AmS}{{\protect\the\textfont2
  A\kern-.1667em\lower.5ex\hbox{M}\kern-.125emS}}

\title{On calculation of the neutron charge radius}

\author{V.A. Karmanov\address{Lebedev Physical Institute,
Leninsky Prospekt 53, 119991 Moscow, Russia}\thanks{e-mails: 
karmanov@sci.lebedev.ru, karmanov@isn.in2p3.fr}}
       
\begin{document}

\maketitle

\begin{abstract}
We show that the anomalous quark magnetic moments and relativistic 
effects in the nucleon wave function result in the correct value of the 
neutron charge radius.
\end{abstract}

\section{INTRODUCTION}\label{sec1}
The neutron charge radius squared $r^2_{En}$ (average of $r^2$ over the 
neutron charge distribution in coordinate space) is expressed through 
derivative of the neutron charge form factor $G^n_E$:
\begin{equation}\label{eq1} 
r^2_{En}=-6\left[\frac{dG_E^n(Q^2)}{dQ^2}\right]_{Q^2=0},
\qquad
G^n_E=F^n_1-\frac{Q^2}{4M^2}F^n_2,
\end{equation}
where $F^n_1,F^n_2$ are the form factors determining the neutron 
electromagnetic vertex:
\begin{equation}\label{ver1} 
\bar{u}'_N\Gamma^{\rho}u_N=
F^n_1(Q^2)\bar{u}'_N\gamma^{\rho}u_N
+\frac{iF^n_2(Q^2)}{2M}\bar{u}'_N\sigma^{\rho\nu}q_{\nu}u_N.
\end{equation} 
Here $u_N,\bar{u}'_N$ are the nucleon spinors, $M$ is the nucleon mass 
and $\sigma^{\rho\nu}=\frac{i}{2}(\gamma^\rho\gamma^\nu - 
\gamma^\nu\gamma^\rho)$.  According to (\ref{eq1}), the neutron charge 
radius is represented as the sum of two contributions:
\begin{equation}\label{eq2} 
r^2_{En}=r_1^2+r^2_{Foldy},
\quad \mbox{where}\quad
r^2_1=-6\left[\frac{dF^n_1(Q^2)}{dQ^2}\right]_{Q^2=0},\quad
r^2_{Foldy}=\frac{3\mu_n}{2M^2},
\end{equation}
$\mu_n=F^n_2(0)$ is the neutron magnetic moment. The contribution 
$r^2_{Foldy}$ is the Foldy term \cite{foldy}, which appears due the 
generation of the electric field by the (anomalous) neutron magnetic 
moment because of its {\it zitterbewegung}.  With the neutron magnetic 
moment $\mu_n=-1.91$ one finds:  $r^2_{Foldy}=-0.126\; fm^2$.  The 
experimental value \cite{kop} $r^2_{En}=-0.113\pm 0.005\; fm^2$ is very 
close to $r^2_{Foldy}$ and corresponds to $r^2_1=0.013\pm 0.005\;fm^2$, 
that is ten times smaller than $r^2_{Foldy}$.  

Most of theoretical calculations, on the contrary, give $r^2_1$ and 
$r^2_{Foldy}$ to be close to each other in absolute values and of 
opposite signs, so, because of cancellation in the sum, $r^2_{En}$ is 
underestimated.  In non-relativistic approximation, with the $SU(6)$ 
symmetric wave function, the neutron charge form factor is identically 
zero (see, e.g., \cite{cs99}): $G^n_E(Q^2)\equiv 0$.  To incorporate 
the relativistic effects, one usually proceeds from this wave function 
and takes into account the relativistic spins rotations by means of the 
Melosh matrices.  They break the $SU(6)$ symmetry \cite{cs99} and give
a non-zero value of $r_{En}^2$.  However, as shown in the paper 
\cite{isgur}, the leading $1/m^2$ correction, calculated in this way 
($m$ is the quark mass), results in $r_1^2=-r^2_{Foldy}=1/(3m^2)$, so, 
in this approximation the value of $r_{En}^2=r_1^2+r^2_{Foldy}$ is 
still zero.  Similar result, by a different method, was found in 
\cite{bc}.  Beyond the $1/m^2$ approximation, $r_1^2$ and $r^2_{Foldy}$ 
don't cancel each other, that can be seen from their analytical 
expressions up to the $1/m^4$:
\begin{equation}\label{eq5}
r_1^2=\frac{1}{3m^2}-\frac{\epsilon}{18m^3}-\frac{7\beta^2}{54m^4},
\;
r_{Foldy}^2=-\frac{1}{3m^2}
+\frac{7\epsilon}{54m^3}-\frac{\beta^2}{27m^4},
\;
r^2_{En}=\frac{2\epsilon}{27m^3}-\frac{\beta^2}{6m^4},
\end{equation}
where $\epsilon=M-3m$, $\beta$ is the parameter of the oscillator wave 
function (see eq.  (\ref{eq53}) below).  We found (\ref{eq5}) by the 
methods of explicitly covariant light-front dynamics 
\cite{cdkm,karm98}.  Since $\epsilon \approx m$ and $\beta/m>1$, one 
can expect large high order corrections and invalidity of the $1/m$ 
decomposition at all.  The numerical calculations without any $1/m^2$ 
approximation \cite{cs99,cpss,Frederico} don't reveal a large effect 
and still underestimate $r^2_{En}$:  $r^2_{En}(theor)\approx (0.65\div 
0.70) r^2_{En}(exp)$. In the paper \cite{wagen} the value 
$r^2_{En}=-0.12\;fm^2$ close to the experimental one was obtained, in 
the point form of dynamics, though there is a deviation of the nucleon 
magnetic moments.

A reliable calculation of the neutron charge radius (as well as of the 
nucleon e.m. form factors) should incorporate many contributions, such 
as relativistic effects in the nucleon wave function, its nontrivial 
spin structure, exchange currents within nucleon, pionic fluctuations 
of constituent quarks \cite{glr}, anomalous quark magnetic moments, 
etc. In this work we will show that the main discrepancy is easily 
removed, when the anomalous quark magnetic moments together with the 
relativistic effects are taken into account. The values of the quark 
magnetic moments are turned out to be close to those found in 
literature from other fits.  Because of approximate and heuristic 
character of our consideration, this does not exclude other 
contributions, but shows importance of those which we take into 
account. 

\section{SCALAR COUPLING AND ANOMALOUS QUARK MAGNETIC\\ MOMENTS}
The full basis for the nucleon wave function contains sixteen spin 
structures \cite{karm98}.  We will take the simplest one:  
$I(12,3)=(\bar{u}_1\gamma_5 U_c\bar{u}_2)(\bar{u}_3 u_N)=-I(21,3)$ 
corresponding to scalar coupling\footnote{Note that the parities of 
$\bar{u}$ and $U_c\bar{u}$ are opposite, hence $\bar{u}_1\gamma_5 
U_c\bar{u}_2$ is a true scalar, just due to $\gamma_5$.}. Here 
$\bar{u}_{1,2,3}$ are the (conjugated) quark spinors.  The same 
coupling (but not only it), to construct the nucleon wave function, was 
used in the paper \cite{Frederico}.  

We represent the symmetric spin-isospin part of the nucleon wave 
function in the form:
\begin{equation}\label{eq50}
\psi=I(12,3)\cd \xi(12,3)+I(23,1) \cd \xi(23,1)
+ I(31,2)\cd\xi(31,2), 
\end{equation} 
where $\xi(12,3)=(\xi_1^\dagger 
i\tau_2\xi_2^\dagger)(\xi_3^\dagger\xi_N)=-\xi_{21,3}$ is the isospin 
part with the zero pair isospin, $\xi$ is an isospinor. This wave 
function has the same non-relativistic limit as the Melosh wave 
function, but it differs from the Melosh one in the relativistic 
domain.  We assume that in neutron $123=ddu$. Calculating (\ref{eq50}) 
for the particular isospinors $\xi{d,d,u}$, we get:  
$\psi_n=-I(23,1)+I(31,2)$.  To simplify the further calculations, it is 
convenient to return to the canonical order 123.  Using the Fierz 
identities, we transform $\psi_n$ to the form:
\begin{equation}\label{eq52}
\psi_n
=-\frac{1}{2}(\gamma^{\mu})_{12}(\gamma_{\mu}\gamma_5)_{3N}+
\frac{1}{4}(\sigma^{\mu\nu})_{12}(\sigma_{\mu\nu}\gamma_5)_{3N},
\end{equation}
where:  $(\gamma^{\mu})_{12}=\bar{u}_1\gamma^{\mu}U_c\bar{u}_2$, 
$(\gamma_{\mu}\gamma_5)_{3N}=\bar{u}_3\gamma_{\mu}\gamma_5 u_N$, etc.  
Note that after Fierz transformation the initial scalar coupling 
(\ref{eq50}) is transformed into the sum of vector and tensor couplings 
in (\ref{eq52}).  The total wave function is the product of the spin 
and the momentum parts:  $\Psi_n=\psi_n\psi_0.$ For $\psi_0$ we chose 
the oscillator form:  
\begin{equation}\label{eq53} \psi_0=N\exp\left(-{\cal 
M}^2/2\beta^2\right), 
\end{equation} 
where ${\cal M}^2=(k_1+k_2+k_3)^2$ is the invariant mass squared of 
$3q$-system, $k_{1,2,3}$ are the on-mass shell quark four-momenta, $N$ 
is found from the normalization condition \cite{karm98}.

Let us introduce the anomalous quark magnetic moments $\kappa_q$, 
representing the quark e.m. vertex similarly to the nucleon one, eq.  
(\ref{ver1}):
\begin{equation}\label{eq57} 
\bar{u'}_q\Gamma^{\rho}u_q= e_q\bar{u'}_q\gamma^{\rho}u_q
+\frac{i\kappa_q}{2m}\bar{u'}_q\sigma^{\rho\nu}q_{\nu}u_q.
\end{equation}
We neglect the dependence of the quark form factors on $Q^2$.  The 
nucleon form factors are calculated in the explicitly covariant 
light-front dynamics \cite{cdkm,karm98}, in the impulse approximation, 
via plus-component of current, by a program of analytical calculation.  
The two-loop (five-dimensional) integrals are calculated numerically.

In the $1/m^2$ order the results can be obtained analytically:
\begin{equation}\label{eq59} 
r_1^2=\frac{1}{6m^2}+\frac{5\kappa_u}{3m^2}+\frac{7\kappa_d}{3m^2},
\quad
r^2_{Foldy}=-\frac{1}{3m^2}-\frac{\kappa_u}{6m^2}
+\frac{2\kappa_d}{3m^2}.
\end{equation}
For the neutron charge radius we thus find:
\begin{equation}\label{eq62}
r_{En}^2=r_1^2+r^2_{Foldy} =-\frac{1}{6m^2}
+\frac{3\kappa_u}{2m^2}+\frac{3\kappa_d}{m^2}.
\end{equation}
Note that when $\kappa_u=\kappa_d=0$, we get 
$r_{En}^2=-\frac{1}{6m^2}$.  So, with the wave function (\ref{eq50}) 
there is no cancellation in the $1/m^2$ order  found in \cite{isgur} 
with the Melosh wave function (we remind that both wave functions 
coincide with each other in non-relativistic limit).  This cancellation 
is a peculiarity of the Melosh wave function.

The fit of the nucleon static properties with $m=340$ MeV, $\beta=700$ 
MeV/c (that corresponds to $\beta^2/m^2\approx 4$, $<k>=420$ MeV/c), by 
numerical calculation not using the $1/m$ decomposition, results in 
following anomalous quark magnetic moments: 
\begin{equation}\label{kappas}
\kappa_u=0.13,\quad \kappa_d=-0.11.
\end{equation}
With these values we get:
\begin{eqnarray*} 
\mu_n=-1.92,\;\mu_p=3.09&\mbox{and}&  r^2_{En}=-0.125\,fm^2\;
(r^2_1=0.002\,fm^2,\; r^2_{Foldy}=-0.127\,fm^2).
\end{eqnarray*}
The values of $\mu_n$ and $r^2_{En}$ are rather close to the 
experimental ones, $\mu_p$ is by 10 \% larger. With the wave function 
(\ref{eq50}), incorporating only one spin structure of sixteen, and 
with a simplest, oscillator parametrization we can hardly pretend to 
a higher accuracy.  Note that the calculation by the approximate 
formula (\ref{eq62}) gives rather close result:  
$r_{En}^2=-\frac{1}{6m^2} 
+\left(\frac{3\kappa_u}{2m^2}+\frac{3\kappa_d}{m^2}\right)= 
-0.056+(-0.045)\approx -0.1 \;fm^2$. From here one can see that half of 
$r^2_{En}$ results from the relativistic effects in the nucleon wave 
function (the term $-\frac{1}{6m^2}$). Another half results from the 
quark magnetic moments. 

For comparison, we give $\kappa_q$'s found in other papers:  
$\kappa_u=0.09,\kappa_d=-0.15$ \cite{cpss}, $\kappa_u=-0.05\div 
0.04,\kappa_d=0.01\div -0.05$ \cite{chung}, 
$\kappa_u=0.36,\kappa_d=-0.18$ \cite{boffi}\footnote{ In ref.  
\cite{boffi}, in contrast to eq.  (\ref{eq57}), the quark charge $e_q$ 
is separated as a factor at the front of both form factors.  The 
original values in \cite{boffi} are:  $\kappa_u=\kappa_d=0.54$.  
Multiplying 0.54 by the quark charges 2/3 for $u$ and $-1/3$ for $d$, 
we obtain the above values of $\kappa_q$'s from \cite{boffi} in the 
definition (\ref{eq57}).}.  Though they differ from each other, these 
values (except for \cite{chung}), as well as the values found in the 
present work, have a common features:  ({\it i}) the absolute values of 
$\kappa_q$'s are of the order of 0.1; ({\it ii}) $\kappa_u$ is 
positive, $\kappa_d$ is negative. These features are enough to 
reproduce the experimental data, whereas the sensitivity of $r^2_{En}$ 
to $\kappa_q$ is high.  Putting $\kappa_q=0$ or changing the sign of 
one or of both $\kappa_q$'s, we obtain the neutron magnetic moment and 
the charge radius incompatible with experiment (see Table 
\ref{table:1}).  
\begin{table}[htb] \caption{Sensitivity of $\mu_n$ and 
$r^2_{En}$ to $\kappa_q$'s.  The best fit: 
$\kappa_u=+0.13,\kappa_d=-0.11$} \label{table:1} 
\begin{tabular}{cc|c|cccc} \hline && $\kappa_u=+0.13$ & 
$\kappa_u=0$ & $\kappa_u=-0.13$ & $\kappa_u=+0.13$ & $\kappa_u=-0.13$ 
\\ &\mbox{Exp.:}& $\kappa_d=-0.11$ & $\kappa_d=0$ & $\kappa_d=-0.11$ & 
$\kappa_d=+0.11$ & $\kappa_d=+0.11$ \\ \hline $\mu_n$& $-1.91$ & 
$-1.92$ & $-1.46$ & $-1.70$ & $-1.21$ & $-1.00$\\ $r^2_{En},fm^2$ 
&$-0.113$ & $-0.125$ & $-0.07$ & $-0.20$ & $+0.07$ & $-0.01$\\ \hline 
\end{tabular} 
\end{table}

We conclude that the anomalous quark magnetic moments (with the values 
close to already known in literature) and the relativistic effects in 
the nucleon wave function easily result in the correct neutron charge 
radius. 

I am grateful to L.Ya. Glozman for stimulating discussions attracting 
my attention to this problem.

\end{document}